# The Benefits of Prosociality towards AI Agents: Examining the Effects of Helping AI Agents on Human Well-Being

Zicheng Zhu
National University of Singapore
Singapore, Singapore
zicheng@u.nus.edu

Yugin Tan
National University of Singapore
Singapore, Singapore
tan.yugin1@gmail.com

Naomi Yamashita
Kyoto University
Kyoto, Japan
naomiy@acm.org

Yi-Chieh Lee
National University of Singapore
Singapore, Singapore
yclee@nus.edu.sg

Renwen Zhang
National University of Singapore
Singapore, Singapore
r.zhang@nus.edu.sg

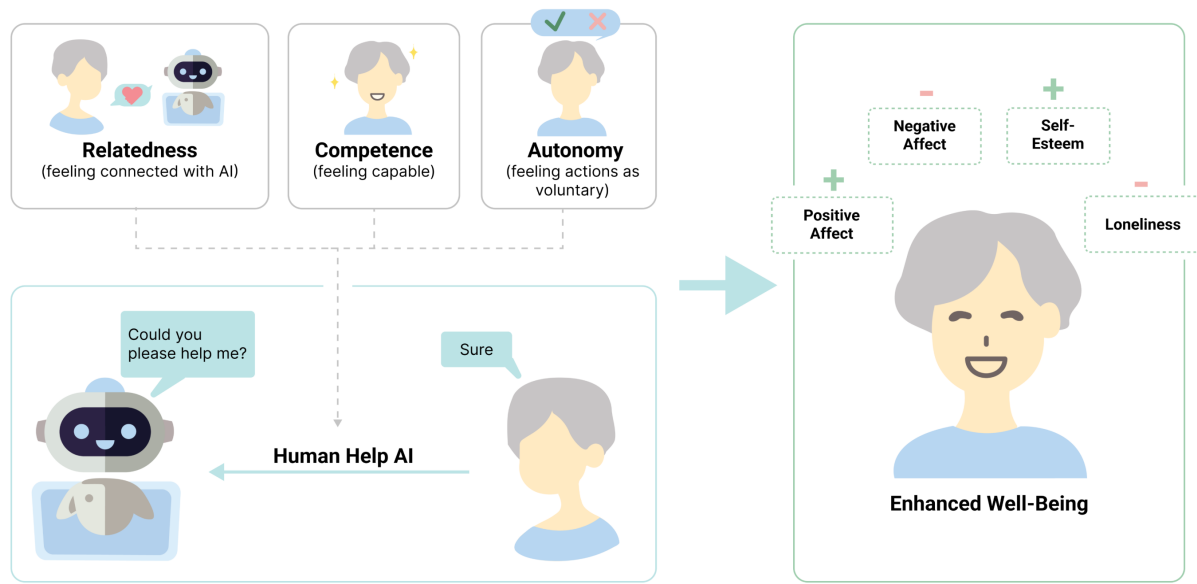

Figure 1: A visual summary of the focus of our study. Our study focuses on how helping AI agents and how satisfying human basic psychological needs—relatedness, competence, and autonomy—by AI during this interaction influences human well-being.

## Abstract
Prosocial behaviors, such as helping others, are well-known to enhance human well-being. While there is a growing trend of humans helping AI agents, it remains unclear whether the well-being benefits of helping others extend to interactions with non-human entities. To address this, we conducted an experiment (N = 295) to explore how helping AI agents impacts human well-being, especially when the agents fulfill human basic psychological needs—relatedness, competence, and autonomy—during the interaction. Our findings showed that helping AI agents reduced participants' feelings of loneliness. When AI met participants' needs for competence and autonomy during the helping process, there was a further decrease in loneliness and an increase in positive affect. However, when AI did *not* meet participants' need for relatedness, participants experienced an increase in positive affect. We discuss the implications of these findings for understanding how AI can support human well-being.

## 1 Introduction

> "For it is in giving that we receive" — Saint Francis of Assisi.

For a long time, there has been great interest in the link between engaging in prosocial behaviors and one's own well-being. Numerous studies have demonstrated that such behaviors produce reliable benefits for human well-being [2, 17, 31, 40, 70]. Such well-being benefits have been observed in people of all ages across global samples [1, 11]. Research has also found that compared to receiving help, offering help leads to more joy [65]. In light of these benefits, prosocial acts have even been used as interventions to improve people's well-being [32, 42].

The well-being benefits of prosocial behaviors have also inspired the design of human-agent interaction. Okada proposed the concept of the "weak robot," which suggests designing robots with imperfections that require human help to enhance human well-being [25]. Nonetheless, this concept has yet to receive empirical support and has primarily focused on embodied agents, while overlooking the more ubiquitous disembodied ones. More importantly, although there has been a growing trend of humans helping disembodied AI agents [19, 27, 33, 46, 64], research in this area has primarily focused on humans helping AI to either improve AI performance [19, 46] or enhance humans' own skills through helping AI agents learn these skills [27, 33]. It remains unclear how helping AI agents impacts human well-being.

Furthermore, previous studies have shown that human well-being is enhanced more when their basic psychological needs—competence (feeling capable), autonomy (perceiving one's actions as voluntary), and relatedness (feeling connected to others)—are satisfied during prosocial behaviors [22, 58]. However, the effects of AI agents fulfilling these needs on human well-being have been unclear, as existing studies primarily focus on how intelligent agents meet these needs to increase user engagement with the agent or a specific task [13, 38, 66]. Considering the potential differences in how people's psychological needs are met by humans and AI agents [7, 19, 30]—such as relatedness being more reciprocal with humans and more self-oriented with AI [7]—there might be differences in the impact on human well-being when these needs are met by AI versus humans. Therefore, it is crucial to examine how satisfying human psychological needs through AI affects their well-being.

In this study, we examine how helping AI agents influences human well-being and, specifically, how satisfying human psychological needs by AI agents during this process affects their well-being. We conducted an online experiment (N = 295) with two main conditions: one where participants helped an AI agent, and another where they did not. Within the help condition, we further assigned participants into groups where the AI agent either satisfied or did not satisfy their basic psychological needs. For each group, we measured participants' well-being before and after their interaction with the agent. Our findings showed that helping AI agents reduced participants' loneliness. Fulfilling participants' needs for competence and autonomy through AI agents during the helping process further decreased participants' loneliness and increased positive affect. However, when the AI agent did *not* meet participants' needs for relatedness, participants displayed a greater increase in positive affect.

Our research presents several contributions to HCI. Firstly, we offer empirical evidence demonstrating that helping AI agents can enhance human well-being. This suggests that, amidst the growing trend of people helping AI, it is important to go beyond just focusing on improving AI performance, but also consider enhancing human well-being during these interactions. Secondly, we offer insights into how AI satisfying human basic psychological needs impacts human well-being. This offers guidance on designing AI to better support human needs and enhance human well-being in scenarios where humans help AI. Thirdly, we offer a novel perspective on AI-aided intervention to improve well-being, as our findings suggest that in addition to using AI to help humans, we can also engage humans to help AI to enhance human well-being.

## 2 Related Work

### 2.1 Enhancing Well-being through Prosocial Behaviors

Prosocial behavior refers to any action intended to benefit others [15]. It is well-known that prosocial behaviors not only benefit the recipient, but also improve the helper's well-being [2, 31, 40, 70]. For instance, Weinstein and Ryan find that helping others complete a task increases one's positive affect and self-esteem and reduces their negative affect [70]. Volunteer work has also been shown to reduce loneliness in elderly people [10]. Furthermore, compared with receiving help, helping others has been found to result in more joy [65]. In light of these benefits, prosocial acts have been utilized as interventions to enhance people's well-being [32, 42]. For instance, people were encouraged to engage in prosocial behaviors during COVID-19 to safeguard their mental health [42]. Preadolescents are also instructed to perform kind acts towards others to boost their well-being [32].

Interestingly, these benefits of prosocial behaviors have also inspired the design of human-agent interaction. Okada proposed the concept of the "weak robot", which suggests designing robots that have imperfections and require human help to improve human well-being [25]. Based on this concept, researchers have developed weak robots like the Sociable Trash Box and iBones, which request human help with tasks like collecting trash and dispensing tissues [50, 75]. It has been observed that people appear happy to help these robots, presumably due to an increased sense of achievement or a stronger connection to the robots [50, 75]. However, these assumptions have not been empirically proven. Moreover, these studies focus primarily on embodied agents and overlook disembodied ones like chatbots, which might be more commonly encountered due to their flexibility and cost-effectiveness. Therefore, there exists a need to explore how helping disembodied AI agents affects human well-being.

## 2.2 The Increasing Trend of Humans Helping AI

In recent years, there has been a growing trend of humans helping AI, driven in part by the increasing need for human input. Currently, human-AI collaboration has evolved from merely delegating tasks to AI into forming human-AI teams where human and AI expertise complement each other [77]. In such collaboration, AI needs human help to compensate for its limitations. For instance, AI requests help from humans to enhance its image recognition ability [19] or to correct biases in its decision-making processes [46]. However, existing research in this area mainly focuses on how to solicit help from humans to contribute to AI [46, 61], with little attention given to the impact of such help on human well-being.

On the other hand, the growing trend of humans helping AI stems from the increasing integration of intelligent agents into daily lives as social actors [49]. As agents are perceived as social actors, people are likely to interact socially and even prosocially with these agents [49]. So far, research in this area has mainly focused on how helping intelligent agents learn specific skills can improve humans' mastery of these skills [27, 33, 64]. For instance, research has explored improving people's mastery of English vocabulary through correcting robots' language errors [64], increasing people's willingness to seek support for depression by allowing them to write supportive messages to a social bot displaying depressive symptoms [27], and enhancing people's self-compassion by enabling them to help AI agents that make mistakes [33]. Despite the insights these studies offer, it remains unclear how providing help to intelligent agents impacts human well-being. Therefore, our aim is to fill this gap by empirically examining the impact of helping AI agents on human well-being. Given the long-established link between prosocial behaviors towards other humans and people's own well-being [2, 31, 40, 70], we propose the following hypothesis:

**H1:** Helping AI agents improves people's well-being, including (a) increased positive affect and (b) increased self-esteem, along with (c) reduced negative affect and (d) reduced loneliness, compared to those who do not help.

## 2.3 The Role of Human Basic Psychological Needs

Besides demonstrating that prosocial behaviors towards others can enhance well-being [2, 31, 40, 70], previous studies suggest that some forms of prosocial behaviors are more beneficial than others [3, 40, 70]. These studies are primarily grounded in Self-Determination Theory (SDT), which asserts that human well-being is enhanced when their basic psychological needs are fulfilled [22, 58]. Specifically, SDT describes three human basic psychological needs: (1) *autonomy*, the need to perceive one's actions as voluntary; (2) *competence*, the need to view oneself as capable; and (3) *relatedness*, the need to feel connected to others [22, 58].

Drawing on SDT, previous research has shown that prosocial behaviors that satisfy human basic psychological needs better enhance well-being [3, 40, 70]. Regarding autonomy, for instance, voluntary rather than compelled help leads to greater increases in positive affect and self-esteem [70]. People also experience more positive emotions when they recall instances of voluntary instead of compelled help [36]. For competence, studies show that making people aware of the positive impact of their donations results in greater happiness [3]. When people reflect on experiences of spending money on others, they also derive more happiness if they can clearly see the impact [36]. Finally, with relatedness, studies have found that spending money on those with whom people have a stronger connection increases people's happiness [4, 36].

Despite these benefits, it is important to note that the well-being impact of AI satisfying human psychological needs when humans help AI may differ from human-to-human interactions due to fundamental differences between humans and AI agents. For instance, although people can develop a sense of relatedness with AI, this relatedness is often perceived as more self-oriented, whereas relatedness with humans is typically seen as more reciprocal [7]. Also, while people may gain a sense of competence from helping AI by seeing improvements in AI accuracy, this help can paradoxically lead them to perceive the AI as less capable [19] and potentially evoke negative emotions. Additionally, since AI agents are generally not viewed as moral entities [30], even if AI affects human autonomy, people may not attribute this influence to the AI, resulting in little or no change in their mood.

Given these potential differences between humans and AI, it is essential to explore how AI satisfying human basic psychological needs impacts human well-being. Nonetheless, existing research on intelligent agents satisfying human basic psychological needs has mainly focused on how this influences people's engagement with the agent or the task [13, 38, 66]. For instance, research has examined the impact of social robots meeting human basic psychological needs on human motivation to communicate with the robots and engage with learning materials [38, 66]. Research has also explored how disembodied agents fulfilling human basic psychological needs improve human engagement [13, 38, 66]. Despite the insights these studies offer, the impact of agents satisfying human basic psychological needs on human well-being remains unclear. Therefore, we aim to examine this issue, guided by the following research questions.

**RQ1:** How do AI agents satisfying human basic psychological needs (competence, autonomy, and relatedness) affect people's well-being when they help these agents?

**RQ2:** What is the interaction effect of AI agents satisfying different psychological needs on people's well-being when they help these agents?

## 3 Method

We conducted a randomized online experiment with two main conditions: one where an AI agent asked for help from participants, and another where the agent did not ask for help. Specifically, within the condition where participants helped the agent, we had sub-conditions where participants' needs for **relatedness**, **competence**, and **autonomy** were met or not met, which resulted in nine groups in our experiment: 2 (relatedness vs. non-relatedness) x 2 (competence vs. non-competence) x 2 (autonomy vs. non-autonomy) + 1 (no help).

## 3.1 Materials

*3.1.1 Mandatory Task.* Following Weinstein and Ryan's work [70], we did not reveal the existence of the helping task to participants but instead informed them that the mandatory task of the experiment was to test a newly developed chat interface and to complete a questionnaire on the interface. This approach aimed to minimize the influence of disclosing the experiment's real purpose on participant behavior.

During the mandatory task, participants entered a chat interface where an AI agent named Chati greeted them, explained the questionnaire task, and informed the participants that an automated system, rather than the agent, would guide them through the questionnaire. This transition to an automated system was to ensure participants did not perceive the act of completing the questionnaire as a form of helping the AI agent.

The questionnaire comprised 20 questions related to the participants' habits of using technology. The questionnaire combined items from existing surveys measuring mobile phone usage [26, 59] with some self-created questions. For the existing survey items, we selected more neutral items (e.g., "I prefer communicating through my mobile phone over face-to-face communication") rather than items measuring problematic mobile phone usage (e.g., "My job and/or school work suffer from the amount of time I spend on the Internet"). This was to avoid reminding participants of negative experiences that might influence their positive or negative affect, which was measured in this study. Similarly, for the self-created questions, we also designed neutral items to minimize any impact on participants' affect (e.g., "I often play games on my mobile phone"). Participants responded on a scale from 1 (strongly disagree) to 7 (strongly agree). The details of these questions can be found in the supplementary material.

*3.1.2 Helping Task.* For the helping task, the agent informed participants that it was developing a messaging app and sought to gain insights from people. Specifically, after participants completed the mandatory task, the automated system redirected participants back to their conversation with the agent. The agent first thanked participants for their questionnaire responses and expressed curiosity about which messaging apps they typically used. After the participant responded, the agent revealed that it was developing a messaging app and inquired about the participants' preferences for color design, showing three images of messaging apps in various colors. This approach was intended to reinforce the participants' belief that the agent was genuinely developing a messaging app. Following the participants' responses, the agent expressed gratitude for their input, stated its desire to gather more insights from the participants, and requested help from them. For participants in the autonomy group, the agent emphasized that this part was completely optional (see details in section 3.1.5).

Once the participants agreed to help, the agent asked for their opinions on the development of the messaging app. The questions included: (1) *"First, I'm curious what you like about your most frequently used messaging app?"* (2) *"I'm also curious if there are any features that you feel are missing or aspects you think could be enhanced?"*; (3) *"I'd also love to get a clearer picture of your daily messaging interactions, such as when and where you text."*; (4) *"Apart from this, I'm also thinking of adding AI features to the messaging app. I'm wondering what tasks you would like AI to help with if this feature were available?"*; and (5) *"And finally, I am curious what would make you consider exploring a new messaging app?"* The agent waited for the participant's response before posing the next question.

We selected this task as the helping task for three reasons: (1) Helping to fulfill others' needs is one of the most common types of prosocial tasks [14], which involves a situation where one party has a specific goal and requires help from another party to achieve it; (2) As AI generative models evolve, they begin to create new content such as images, music, and even applications. In this process, having a human-in-the-loop is crucial. Humans can step in, provide feedback, and help the algorithm improve; (3) The helping task should be neither too challenging nor too specialized to avoid situations where participants choose not to help simply because they feel incapable. Considering that over 5 billion people use messaging apps as of 2024 [16], we decided to solicit participants' input on messaging app development as our helping task, which ensured that they were capable of contributing.

*3.1.3 Manipulation of Relatedness.* To foster relatedness between the agent and participants in the relatedness group, we followed Walton et al.'s approach [68] to initiate a casual conversation between the agent and participants. Specifically, the conversation was about music and movies. As sharing similar preferences has been found to increase relatedness between individuals [68], we designed the agent to occasionally express similar preferences to those of the participants to increase relatedness. For instance, if a participant said, *"I like Taylor Swift because of her great genius, her innovations, and her brilliance as a songwriter, both melodically and verbally,"* the agent responded, *"No way, Taylor Swift? Huge fan here too! 'All Too Well' just hits differently — her storytelling is unmatched."* To avoid participants' perception of the agent merely pandering to them, the agent also expressed different preferences on certain topics. This casual conversation lasted approximately 5 minutes.

For participants assigned to the "non-relatedness" and "no help" groups, they did not engage in casual conversation with the agent, as such interactions could inadvertently enhance participants' feelings of relatedness with the agent. Notably, skipping this segment would cause a difference in experiment duration between the "non-relatedness" and "no help" groups and the "relatedness" group. Such differences in duration could introduce unwanted effects on the dependent variables [44]. For example, participants in the "relatedness" group who engaged in casual conversation with the agent might experience greater fatigue due to the longer experiment, potentially leading to lower measured well-being at the end compared to those in the "non-relatedness" group. To avoid this confounding effect, we followed a previous design [71] and had participants in the "non-relatedness" and "no help" groups complete a filler task during the same 5-minute period allocated for the casual conversation in the "relatedness" group. Specifically, the "non-relatedness" and "no help" groups completed a questionnaire consisting of 16 questions related to their technology usage, such as *"I use technology primarily for entertainment purposes,"* with options ranging from 1 (strongly disagree) to 7 (strongly agree). Similar to the approach used in Sect 3.1.1, these questions were a combination of items from existing surveys on mobile phone usage [26, 59] and questions we created. Notably, none of these items overlapped with those used in the mandatory task. The details of these questions are available in the supplementary material. This design can be viewed in Figure 2.

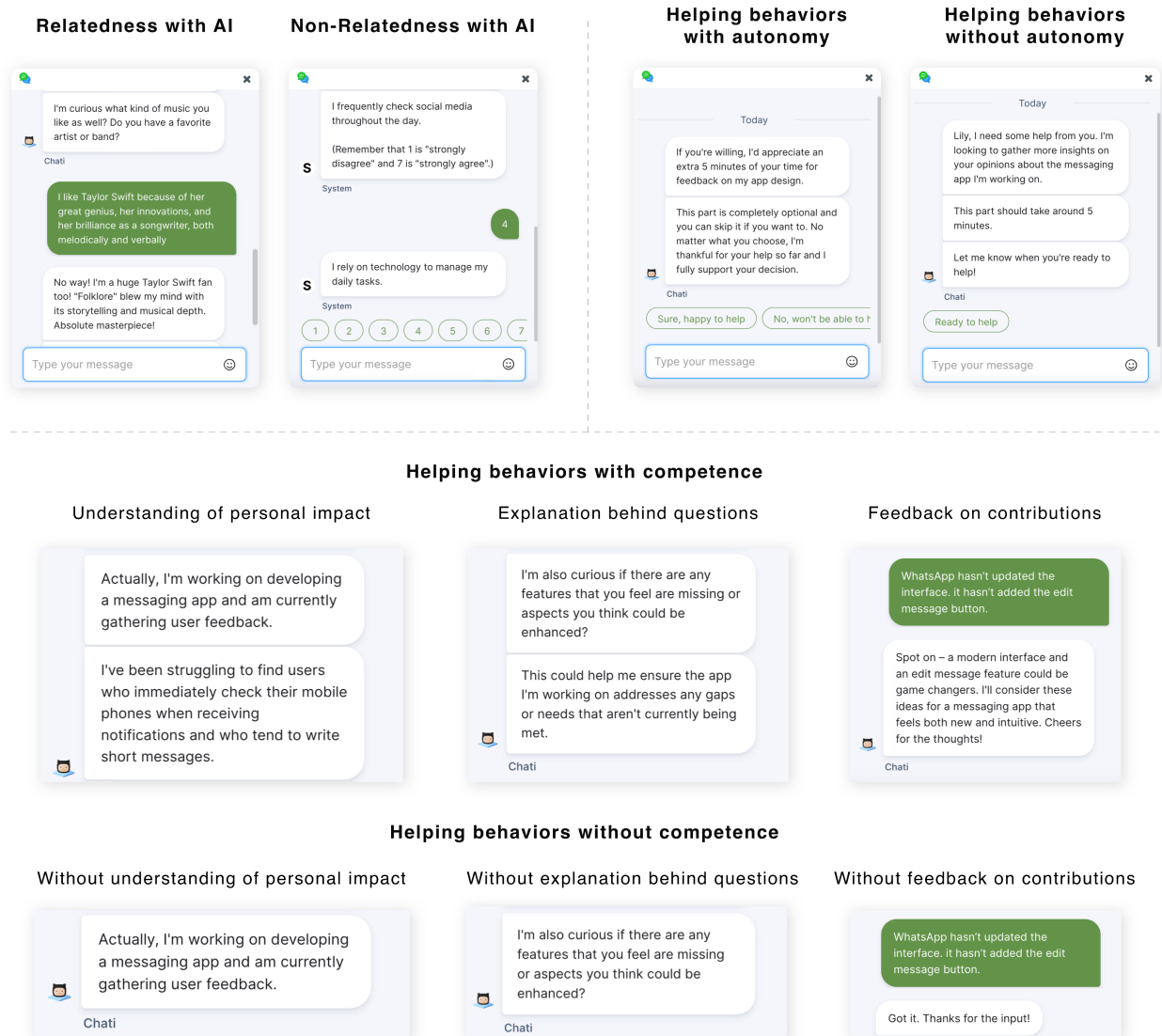

Figure 2: Agent manipulations for study conditions with and without relatedness (top left), autonomy (top right), and competence (bottom). For relatedness, participants in the relatedness group engaged in casual conversation with the AI agent, while those in the non-relatedness group did not. For autonomy, participants in the autonomy group could choose whether to help the agent, whereas those in the non-autonomy group were required to help. For competence, the AI agent in the competence group explained the personal impact of the participant's help, provided explanations for its questions, and offered feedback after receiving help, while these elements were absent in the non-competence group.

*3.1.4 Manipulation of Competence.* We manipulated competence based on the definition provided by Aknin and Whillans [5], who described competence in the context of prosocial behaviors as giving prosocial actors clear evidence of how their actions have positively influenced others. Based on this definition, we enhanced competence of participants in the competence group in three aspects (Figure 2):

(1) **Understanding of Personal Impact:** The agent informed participants about their crucial role in the development of

the messaging app. Specifically, the agent mentioned that it had reviewed participants' questionnaire responses and highlighted, *"Reading through your answers, I find that you (do not immediately check/immediately check) your mobile phone when receiving notifications and you tend to write (short/long) messages. I've been struggling to find users who (do not immediately check/immediately check) their mobile phones when receiving notifications and who tend to write (short/long) messages."* The content in parentheses was customized based on the participants' responses in the questionnaire.

(2) **Explanation Behind Questions:** When the agent posed questions, it explained the reason for these questions. For example, after asking, *"I'm curious what you like about your most frequently used messaging app?"* the agent added, *"This can help me understand what makes certain apps popular, which will be useful for my own designs."*

(3) **Feedback on Contributions:** After receiving participants' responses, the agent expressed how their input could influence its work. For instance, when a participant responded, *"I like the simple messaging design. Whatsapp specifically I enjoy the emojis and gif selection. Both messaging and Whatsapp have the capability of video streaming which is great!"* the agent replied, *"Absolutely, video streaming adds so much to chats! I'll consider integrating seamless video features that are user-friendly and fun, just like the emojis and GIFs you love. Thanks for sharing!"* In contrast, for participants in the non-competence group, the agent merely expressed gratitude, such as, *"Thanks for sharing your thoughts!"*

*3.1.5 Manipulation of Autonomy.* Weinstein and Ryan define autonomy in prosocial behavior as when helpers experience a sense of personal choice or volition in providing help [70]. Therefore, following the method they used to manipulate autonomy in prosocial behaviors [70], we manipulated autonomy by allowing participants in the autonomy group to freely choose whether to help or not, while participants in the non-autonomy group had no choice and were required to help the AI agent. Specifically, for participants in the autonomy group, the agent stated, *"This part is completely optional and you can skip it if you want to. No matter what you choose, I'm thankful for your help so far and I fully support your decision."* For participants in the non-autonomy group, the agent underscored both the helping nature of the task and its forced aspect by mentioning, *"I need some help from you. I'm looking to gather more insights on your opinions about the messaging app I'm working on. Let me know when you're ready to help!"* The illustration of these designs can be viewed in Figure 2.

*3.1.6 Agent Implementation.* We developed the chat interface and the agent through UChat[1]. The agent's responses were generated through a combination of a rule-based system and the OpenAI GPT-4 API. Specifically, for interactions that did not require tailored questions or responses, we pre-set the questions and responses. For questions and responses that needed to be customized to the participant's input, we crafted prompts to GPT-4.

Our prompts typically consisted of four parts: context, task, example, and note. The context provided GPT-4 with background information about the conversation between the participant and the agent. The task section instructed GPT-4 on the type of response required. For instance, when asking for "feedback on contributions" in the manipulation of competence, we wrote, *"Express validation of the user's response and appraise for the useful response. You should also let the user know what you will do while designing your message app based on the user's response. You need to provide specific details and examples of the app design you're thinking of based on the user's response. You should also let the user know you will pay attention to this while designing the app."* The example section provided a model answer, and the note section included instructions like *"Keep the response under 30 words"* to make the dialogue resemble a natural conversation. The details of the prompts can be found in the supplementary material.

## 3.2 Procedure

Figure 3 illustrates the procedure of our study. Upon entering the experiment, participants were shown a welcome and consent page and then randomly assigned to one of nine groups. Participants in the relatedness group engaged in a casual conversation with the AI agent to foster relatedness with the agent, while those in the non-relatedness group completed a survey on their technology usage, as described in section 3.1.3. All participants then filled out a survey to measure their baseline well-being before moving on to the mandatory and helping tasks. Following the helping task, participants completed a final survey that included measurements of well-being, a manipulation check, and demographic questions. Participants in the "no help" group skipped the helping task and proceeded directly to the final survey.

Notably, we manipulated relatedness between participants and the agent before the helping task because we aimed to examine whether there was a difference between participants who helped the AI with whom they had already developed relatedness and those who had not. As such, we needed to establish relatedness between the participants and the AI before participants helped the AI. The reason why we further placed relatedness manipulation before the mandatory task was because of the close connection between the mandatory task and the helping task—where the AI agent tailored its help request based on participants' responses in the mandatory task (see details in section 3.1.4). Additionally, the reason why we measured participants' well-being after the relatedness manipulation was to detect any differences in baseline well-being among the groups, since the relatedness manipulation might affect well-being differently for the relatedness and non-relatedness groups. Identifying these initial differences enabled us to account for them as control variables in further analysis.

We conducted the experiment on Qualtrics and we embedded the chat window within Qualtrics. The chat window only appeared when we required participants to enter the chat interface.

## 3.3 Participants

We recruited participants through CloudResearch. A power analysis (power = .80, significance level = .05, effect size f = .25, number of groups = 9) indicated that a minimum of 252 participants was needed, with at least 28 participants in each group. Therefore, we aimed to recruit a minimum of 28 participants per group. We obtained approval from our Ethics Review Committee. The experiment lasted approximately 15-20 minutes. This timing was based on prior research [70], which shows that such brief prosocial acts are sufficient to impact people's well-being. We compensated each participant with USD $3 for their time. For eligibility criteria, participants must be at least 18 years old. Following Weinstein and Ryan's work [70], we dropped participants who chose not to help when requested help (N = 93) from the analysis. We further excluded participants who failed attention checks. There were four attention check questions in the survey, such as "I select agree as I am attentive." Participants who selected the wrong option were excluded, resulting in a total of 48 exclusions. Consequently, our final valid dataset consisted of 295 participants, with each group containing approximately 30-40 participants.

---
[1] https://uchat.au/

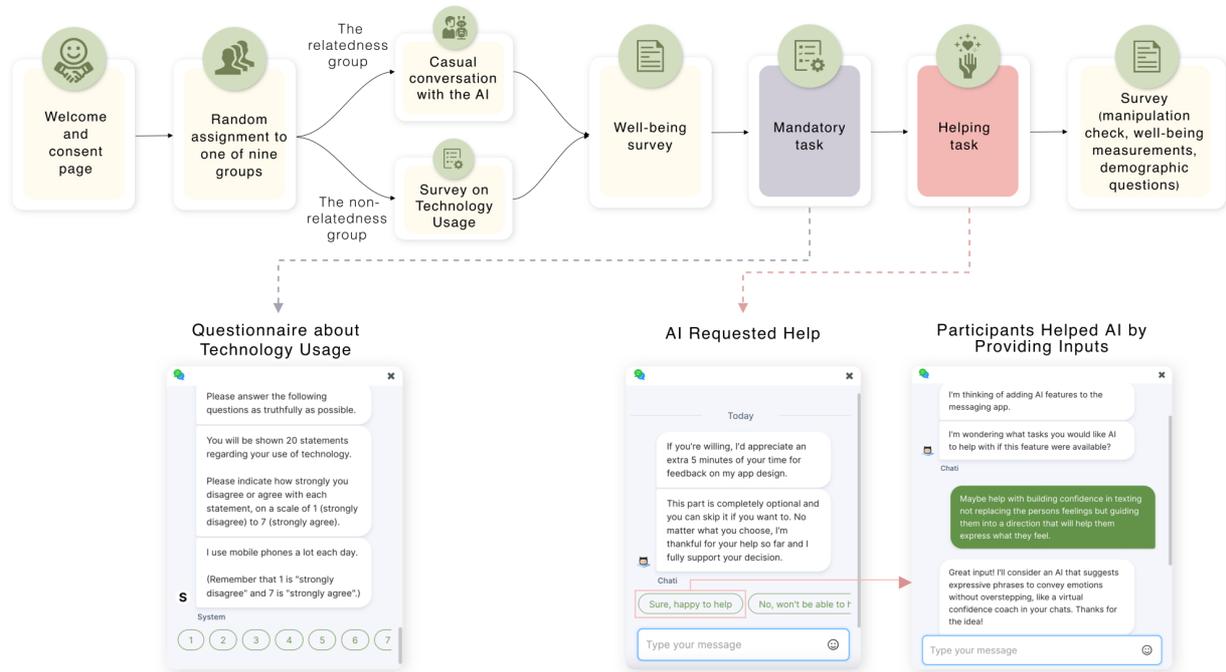

Figure 3: Experiment flow: Participants first viewed a welcome and consent page. They were then randomly assigned to one of nine groups and completed the relatedness manipulation. Next, they filled out a baseline well-being survey, followed by the mandatory task, the helping task, and finally, the manipulation check, post well-being survey, and demographic questions. Participants in the "no help" group skipped the helping task and proceeded directly to the final step.

Overall, the participants had an average age of 38 years (SD = 12.24). The majority held a bachelor's degree (n = 132). 142 were females (48.14%), 152 were males (51.53%), and one preferred not to specify their gender. Detailed participant demographics for each group can be found in Table 1.

### 3.4 Measurement

We measured participants' positive and negative affect, self-esteem, and loneliness as indicators of well-being, given that a large body of work has demonstrated that prosocial behaviors towards other humans impact these aspects of well-being [2, 31, 40, 70]. We measured participants' well-being both at baseline and post-interaction. We also conducted manipulation checks. All measures were rated on a 7-point Likert scale, ranging from 1 (strongly disagree) to 7 (strongly agree), unless otherwise stated. Details of all measurement items are provided in the supplementary material.

**Positive and Negative Affect.** This was measured using the PANAS scale [69]. Positive affect was measured with ten items, such as *"Interested,"* *"Excited,"* and *"Inspired"* (Baseline: $\alpha$ = 0.93; Post: $\alpha$ = 0.95). Negative affect was measured with ten items, such as *"Upset,"* *"Distressed,"* and *"Irritable"* (Baseline: $\alpha$ = 0.94; Post: $\alpha$ = 0.93).

**Self-Esteem.** This was measured using the Rosenberg self-esteem scale [57]. Participants responded to ten statements measuring their current feelings. Sample items include *"I am satisfied with myself"* (Baseline: $\alpha$ = 0.96; Post: $\alpha$ = 0.96).

**Loneliness**. This was measured using three items from the scale developed by Hughes et al. [23], which capture participants' current feelings. Example items include *"I feel that I lack companionship"* (Baseline: $\alpha$ = 0.93; Post: $\alpha$ = 0.96).

**Manipulation Check**. We conducted a manipulation check to confirm the effectiveness of our manipulations regarding relatedness, competence, and autonomy. Specifically, we used a scale adapted from Wilson et al.'s work [73]. We used six items to measure relatedness, such as *"I feel connected to Chati"* ($\alpha$ = 0.92); four items to measure competence, such as *"I am aware of how my responses helped Chati design the app"* ($\alpha$ = 0.97); and four items to measure autonomy, such as *"I was free to choose whether to help with Chati's app development"* ($\alpha$ = 0.98).

Table 1: Participant demographics for each group. In the 'Gender' column, 'F' stands for female, 'M' for male, and 'NA' indicates gender not specified. In the 'Age' column, 'M' represents the mean, and 'SD' represents the standard deviation.

| Group | | | Sample Size | Gender | Age |
| --- | --- | --- | --- | --- | --- |
| Relatedness | Competence | Autonomy | | | |
| with | with | with | 35 | 15F, 20M | 18 - 69 (M = 36, SD = 12.86) |
| without | with | with | 30 | 18F, 12M | 18 - 63 (M = 40, SD = 11.66) |
| without | without | with | 34 | 16F, 18M | 20 - 60 (M = 38, SD = 11.20) |
| without | without | without | 36 | 22F, 14M | 21 - 77 (M = 38, SD = 12.93) |
| with | without | with | 32 | 15F, 17M | 22 - 64 (M = 41, SD = 13.18) |
| with | without | without | 30 | 5F, 25M | 19 - 62 (M = 35, SD = 11.46) |
| with | with | without | 31 | 20F, 11M | 23 - 68 (M = 41, SD = 12.07) |
| without | with | without | 31 | 13F, 17M, 1 NA | 20 - 70 (M = 34, SD = 11.20) |
| no help (control group) | | | 36 | 18F, 18M | 19 - 73 (M = 36, SD = 12.54) |

### 3.5 Analysis

We subtracted participants' baseline well-being scores from the post scores and took this change in well-being as our dependent variable. This approach allowed us to understand the extent and direction of the changes and to compare the differences between groups.

To check the assumptions necessary for standard parametric analysis, we performed a *Levene* test to check the homogeneity of variance and a *Shapiro–Wilk* test to check normality. The Levene test confirmed homogeneity across groups, whereas the *Shapiro–Wilk* test indicated that the normality assumption was not met. Consequently, we employed non-parametric analyses, specifically the *Mann-Whitney U* test or the *Kruskal-Wallis* test followed by a post-hoc *Dunn* test with *Bonferroni* correction. For analyses where non-parametric tests were challenging, such as those requiring the inclusion of control variables, we transformed the dependent variable by taking its logarithmic value, and then proceeded with parametric testing, including t-tests and ANCOVA with post-hoc analyses using *Bonferroni* corrections.

Additionally, we examined potential differences in gender, age, education, and baseline well-being across groups. Where such differences existed, we controlled for these variables in analyses to mitigate any confounding effects.

## 4 Results

### 4.1 Manipulation Check

The Mann-Whitney U Test showed that participants in the relatedness group perceived significantly higher relatedness with the agent (M = 4.94, SD = 1.14) compared to those in the non-relatedness group (M = 3.60, SD = 1.28; W = 16716, $p$ < 0.001). The same was true for competence (M = 5.94, SD = 0.88) compared to non-competence (M = 4.60, SD = 1.45; W = 13084, $p$ < 0.001) and autonomy (M = 6.51, SD = 0.58) compared to non-autonomy (M = 4.68, SD = 1.77; W = 13816, $p$ < 0.001). These results confirmed that our manipulations were successful.

### 4.2 Helping AI Reduces People's Loneliness (H1)

The Mann-Whitney U test revealed that participants who helped the agent (M = -0.47, SD = 0.88) experienced a more significant decrease in loneliness compared to those who did not help (M = -0.16, SD = 0.60; W = 5451, $p$ < 0.05). There were no significant differences in the change in positive affect (Helped: M = -0.00, SD = 0.76 vs. Didn't help: M = -0.33, SD = 0.80; W = 4840, $p$ = 0.10), change in negative affect (Helped: M = -0.09, SD = 0.39 vs. Didn't help: M = -0.07, SD = 0.21; W = 4840, $p$ = 0.68), and change in self-esteem (Helped: M = 0.04, SD = 0.35 vs. Didn't help: M = -0.02, SD = 0.30; W = 5111, $p$ = 0.35). These results can be viewed in Figure 4. Taken together, these results showed that participants who helped the AI agent had a greater decrease in loneliness compared to those who did not help. These results supported hypothesis H1(d) but did not support hypotheses H1(a), H1(b), or H1(c).

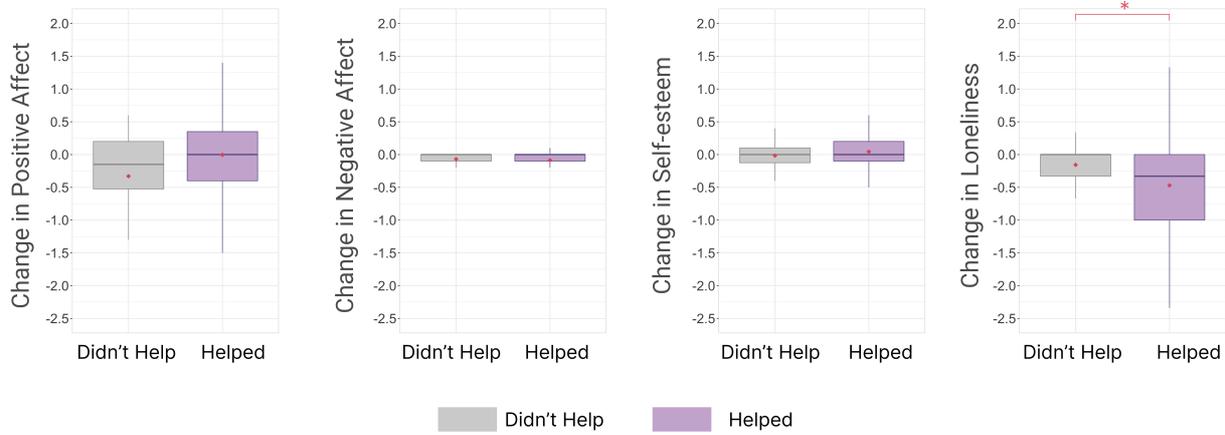

Figure 4: Change in positive affect, negative affect, self-esteem, and loneliness in participants who helped vs. did not help the agent. The results showed that participants who helped the agent had a greater decrease in loneliness compared to those who did not. Whisker lines show the data range from minimum to maximum values. The boxes extend from the 25th to the 75th percentile. The line within each box indicates the median and the red dot indicates the mean. On the y-axis, positive numbers indicate an increase in the variable and negative numbers indicate a decrease. *$p$ < 0.05.

## 4.3 The Impact of AI Satisfying Basic Psychological Needs on Well-being When People Help AI (RQ1)

For **competence**, the Kruskal-Wallis test revealed significant differences among the competence, non-competence, and no help groups in changes in positive affect (H(2) = 16.84, $p$ < 0.001), negative affect (H(2) = 7.16, $p$ < 0.05), and loneliness (H(2) = 13.84, $p$ < 0.001).

For positive affect, the competence group (M = 0.17, SD = 0.80) showed a significantly greater increase compared to both the non-competence group (M = -0.18, SD = 0.68; Z = -3.76, $p$ < 0.001) and the no help group (M = -0.33, SD = 0.80; Z = -2.81, $p$ < 0.05), with no significant difference found between the non-competence and no help groups (Z = -0.34, $p$ = 1.00).

In terms of negative affect, the competence group (M = -0.17, SD = 0.80) experienced a significantly greater decrease than the non-competence group (M = -0.01, SD = 0.68; Z = 2.64, $p$ < 0.05), with no significant difference found between the competence and the no help group (M = -0.07, SD = 0.21; Z = 0.50, $p$ = 1.00) and between the non-competence and the no help group (Z = -1.24, $p$ = 0.64).

For changes in loneliness, the competence group (M = -0.63, SD = 0.89) saw a significantly greater decrease than the non-competence group (M = -0.32, SD = 0.84; Z = 2.74, $p$ < 0.05) and the no help group (M = -0.16, SD = 0.60; Z = 3.29, $p$ < 0.001), with no significant difference between the non-competence and no help groups (Z = 1.50, $p$ = 0.40). There were no significant differences in changes in self-esteem (H(2) = 1.16, $p$ = 0.31) among the competence, non-competence, and no help groups.

These results can be seen in Figure 5 and Table 2. Collectively, these results suggested that meeting participants' needs for competence when they helped the AI agent resulted in a greater increase in positive affect and decrease in loneliness, compared to when these needs were not met or when participants did not help the agent. Meeting participants' need for competence when they helped the AI agent also led to a greater decrease in negative affect compared to when this need was not met.

For **autonomy**, the Kruskal-Wallis test showed significant differences among the autonomy, non-autonomy, and no help groups in change in positive affect (H(2) = 19.07, $p$ < 0.001), change in self-esteem (H(2) = 8.67, $p$ < 0.05), and change in loneliness (H(2) = 17.44, $p$ < 0.001). Specifically, the autonomy group reported significantly greater increases in positive affect (M = 0.18, SD = 0.73) compared to the non-autonomy (M = -0.20, SD = 0.74; Z = -4.04, $p$ < 0.001) and no help groups (M = -0.33, SD = 0.80; Z = -2.88, $p$ < 0.05), with no significant differences noted between the non-autonomy and no help groups (Z = -0.21, $p$ = 0.84).

For self-esteem, the autonomy group saw a significant increase (M = 0.11, SD = 0.34) compared to the non-autonomy group (M = -0.02, SD = 0.35; Z = -2.80, $p$ < 0.05), with no significant differences between the autonomy and the no help groups (Z = -1.80, $p$ = 0.22) and between the non-autonomy and no help groups (Z = 0.04, $p$ = 1.00).

Regarding changes in loneliness, the autonomy group experienced a significantly greater decrease (M = -0.62, SD = 0.83) compared to the non-autonomy group (M = -0.32, SD = 0.91; Z = 3.33, $p$ < 0.01) and no help group (M = -0.16, SD = 0.60; Z = 3.50, $p$ < 0.01), with no significant differences observed between the non-autonomy and no help groups (Z = 1.27, $p$ = 0.62). Lastly, there were no significant differences in the changes in negative affect (H(2)

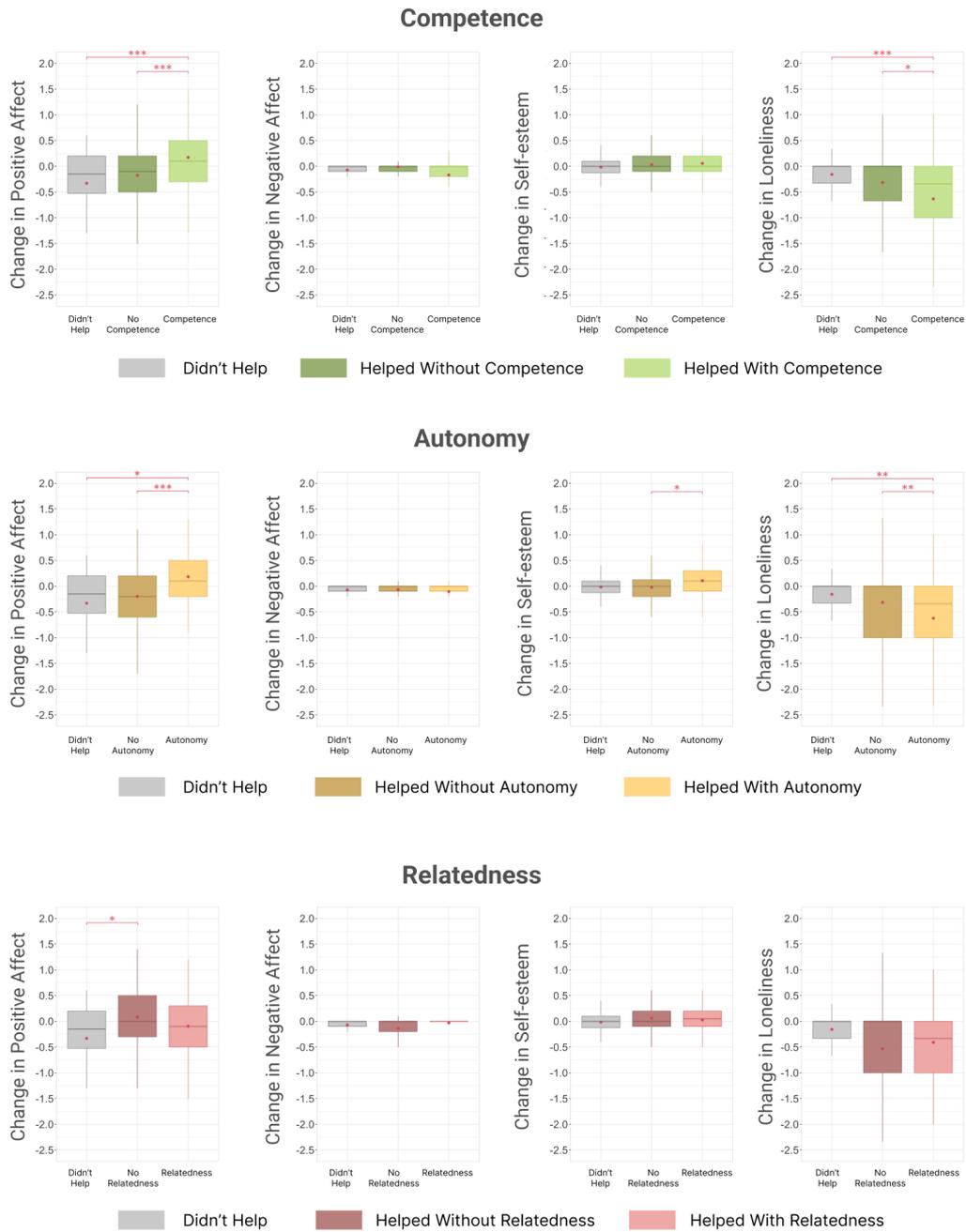

Figure 5: Changes in positive affect, negative affect, self-esteem, and loneliness when competence, autonomy, and relatedness were satisfied vs. not satisfied, with participants who did not help the AI as a control group. The line within each box indicates the median and the red dot the mean. On the y-axis, positive numbers indicate an increase in the variable and negative numbers indicate a decrease. $^{*}p < 0.05$. $^{**}p < 0.01$. $^{***}p < 0.001$.

= 4.21, $p$ = 0.12) among the groups. These results can be seen in Figure 5 and Table 2.

Taken together, these results suggested that meeting participants' needs for autonomy while they helped the AI agent resulted in a greater increase in positive affect and decrease in loneliness, compared to when these needs were not met or when they didn't help the AI agent. Meeting participants' needs for autonomy while they helped the AI agent also resulted in a greater increase in self-esteem compared with when these needs were not met.

For **relatedness**, since there were significant differences or near-significant differences in baseline positive affect (H(2) = 20.15, $p$ < 0.001), negative affect (H(2) = 5.54, $p$ = 0.06), and loneliness (H(2) = 5.55, $p$ = 0.06) among the relatedness, non-relatedness, and no help groups, we included these variables as control variables in our analysis.

The ANCOVA showed significant differences among groups in change in positive affect (F(2, 291) = 4.50, $p$ < 0.05), negative affect (F(2, 291) = 3.15, $p$ < 0.05), and loneliness (F(2, 291) = 3.07, $p$ < 0.05). Post-hoc tests indicated that, for positive affect, the non-relatedness group exhibited a significantly greater increase (M = 0.08, SD = 0.80) compared with the no help group (M = -0.33, SD = 0.80; t = 2.86, $p$ < 0.01), but there were no significant differences between the relatedness and no help group (t = -1.81, $p$ = 0.07), nor between the relatedness and non-relatedness groups (t = 1.50, $p$ = 0.13).

In terms of negative affect, post-hoc test indicated that there was no significant difference between the non-relatedness and the relatedness group (t = 0.98, $p$ = 0.99), between the relatedness and the no help groups (t = -0.12, $p$ = 1.00), and the non-relatedness and the relatedness groups (t = 0.10, $p$ = 0.29). Furthermore, there were no significant differences in changes in self-esteem (H(2) = 0.97, $p$ = 0.62) across the groups.

In terms of loneliness, post-hoc test indicated that the non-relatedness group (M = -0.53, SD = 0.95) approached a significantly greater decrease in loneliness than the no help group (M = -0.16, SD = 0.60; t = 2.38, $p$ = 0.05), with no significant difference between the non-relatedness and the relatedness group (t = -1.24, $p$ = 0.65), and between the relatedness and the no help groups (t = 1.55, $p$ = 0.36).

These results can be seen in Figure 5 and Table 2. Taken together, these results suggested that failing to meet participants' needs for relatedness while they helped the AI agent resulted in a greater increase in positive affect and loneliness compared with when they did not help.

### 4.4 The Interaction Effects of Satisfying Basic Psychological Needs on Well-being When People Help AI (RQ2)

Given differences among groups in baseline positive affect, we included participants' baseline positive affect as a control variable when analyzing the change in positive affect. The three-way ANCOVA indicated a significant interaction effect between competence, autonomy, and relatedness on changes in positive affect (F(1, 250) = 5.73, $p$ < 0.05). This interaction is illustrated in Figure 6. There was no significant interaction effect on changes in negative affect (F(1, 250) = 2.13, $p$ = 0.15), changes in self-esteem (F(1, 250) = 0.14, $p$ = 0.71) or changes in loneliness (F(1, 250) = 0.05, $p$ = 0.83).

Additionally, there was no significant interaction effect between any two variables on participants' well-being. Detailed results are provided in the supplementary materials.

Collectively, these results suggested that when participants lacked relatedness with the AI agent, satisfying their needs for competence and autonomy while helping the AI led to a greater increase in positive affect compared with when they had relatedness with the AI. In contrast, when participants had relatedness with the AI agent, not satisfying their needs for competence and autonomy while helping the AI led to a greater decrease in positive affect compared with when they lacked relatedness with the AI. The summary of all results can be seen in Table 3.

## 5 Discussion
### 5.1 The Well-Being Benefits of Helping AI Agents

Our study uncovered that participants helping AI agents had a greater decrease in loneliness compared with those who did not help. Furthermore, when participants' needs for autonomy and competence were satisfied by the AI agent when they helped the agent, they had a greater increase in positive affect compared to those who did not help. These results, on the one hand, echo previous research showing that prosocial behaviors towards others enhance people's well-being [2, 31, 40, 70]. On the other hand, these findings demonstrate that such well-being benefits observed in helping humans also extend to helping AI agents.

Furthermore, these findings extend beyond the focus of existing research on humans helping intelligent agents, which has primarily focused on how humans can help intelligent agents to improve the agents' performance or to facilitate human skill acquisition [46, 64]. Building on this, our findings suggest that attention can also be paid to the impact on human well-being, no matter whether humans help agents to improve agents' performance or to learn a skill. Specifically, as human-AI collaboration evolves towards partnerships that achieve complementary performance [77], there will be numerous opportunities for people to leverage their expertise to identify AI limitations and help AI. During these interactions, it is important for stakeholders to not just focus on AI performance or team productivity, but also consider the benefits to human well-being, in order to create more meaningful and beneficial human-AI interactions.

Additionally, by uncovering the benefits of helping AI agents on people's well-being, our study provides a new perspective for utilizing AI for human well-being. Previous research on AI for well-being has primarily focused on how AI can offer help to humans, such as providing social support [41], monitoring workers' well-being [12], and offering sleep advice [24]. Our findings add to this line of research by suggesting a reverse direction—enhancing human well-being through humans offering help to AI agents. This approach might be especially beneficial for groups with fewer opportunities to help others, such as the elderly and disabled, who may be perceived as lacking ability [35, 62], or individuals who lack expertise, as people tend to seek help from those who have expertise [45]. As highlighted in prior work, an important means for these individuals to engage in social interaction and integration is to provide them with opportunities to give, not just receive, help [35]. Therefore, we suggest stakeholders consider designing AI agents that require help from these groups of people to engage them in helping behaviors, which may enhance their sense of self-worth and well-being.

Table 2: The mean and standard deviation for each group regarding the change in positive affect, negative affect, self-esteem, and loneliness, along with Bonferroni-corrected p-values for intergroup differences. The numbers under each group represent the mean followed by the standard deviation in parentheses. *$p < 0.05$. **$p < 0.01$. ***$p < 0.001$.

| Variables | Group | | | Bonferroni p value | | |
|---|---|---|---|---|---|---|
| | No Help (NH) | No Competence (NC) | Competence (C) | NH & NC | NH & C | NC & C |
| Positive Affect | -0.33 (0.80) | -0.18 (0.68) | 0.17 (0.80) | 1.00 | 0.05 | < 0.001*** |
| Negative Affect | -0.07 (0.21) | -0.01 (0.68) | -0.17 (0.80) | 0.64 | 1.00 | < 0.05* |
| Self-esteem | -0.02 (0.30) | 0.03 (0.34) | 0.06 (0.37) | 0.47 | 0.29 | 0.60 |
| Loneliness | -0.16 (0.60) | -0.32 (0.84) | -0.63 (0.89) | 0.40 | < 0.01** | < 0.05* |
| Variables | Group | | | Bonferroni p value | | |
| | No Help (NH) | No Autonomy (NA) | Autonomy (A) | NH & NA | NH & A | NA & A |
| Positive Affect | -0.33 (0.80) | -0.20 (0.74) | 0.18 (0.73) | 1.00 | 0.05 | < 0.001*** |
| Negative Affect | -0.07 (0.21) | -0.07 (0.44) | -0.11 (0.32) | 1.00 | 1.00 | 0.5 |
| Self-esteem | 0.30 (-0.02) | -0.02 (0.35) | 0.11 (0.34) | 1.00 | 0.22 | < 0.05* |
| Loneliness | -0.16 (0.60) | -0.32 (0.91) | -0.62 (0.83) | 0.62 | < 0.01** | < 0.01** |
| Variables | Group | | | Bonferroni p value | | |
| | No Help (NH) | No Relatedness (NR) | Relatedness (R) | NH & NR | NH & R | NR & R |
| Positive Affect | -0.33 (0.80) | 0.08 (0.80) | -0.09 (0.71) | < 0.05* | 0.21 | 0.4 |
| Negative Affect | -0.07 (0.21) | -0.14 (0.30) | -0.03 (0.45) | 0.99 | 1.00 | 0.29 |
| Self-esteem | -0.02 (0.30) | 0.06 (0.37) | 0.03 (0.34) | 1.00 | 1.00 | 0.97 |
| Loneliness | -0.16 (0.60) | -0.53 (0.95) | -0.41 (0.81) | 0.05 | 0.36 | 0.65 |

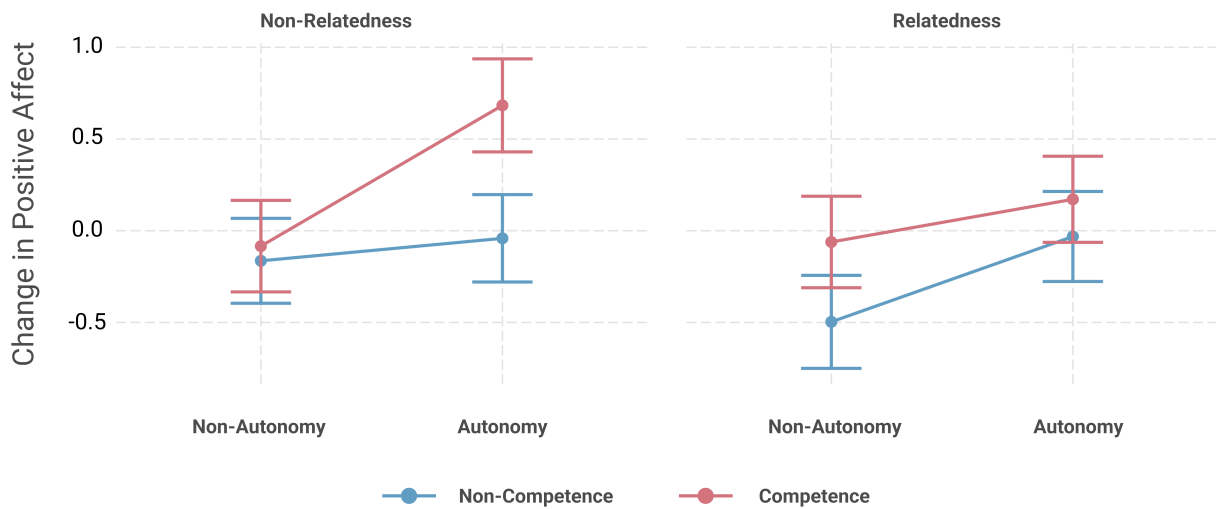

Figure 6: The three-way interaction effects between autonomy, competence, and relatedness satisfaction on participants' change in positive affect when they help the AI agent. The results showed that when participants lacked relatedness with the AI, satisfying their needs for competence and autonomy led to a greater increase in positive affect compared with when participants had relatedness with AI. When participants had relatedness with the AI, not satisfying their needs for competence and autonomy resulted in a greater decrease in positive affect compared with when participants did not have relatedness with AI.

Our study found that even brief acts of helping an agent improved participants' well-being. Although we did not measure changes in participants' well-being over time, previous studies find that engaging in long-term prosocial behavior can lead to lasting improvements in well-being [18, 42]. For example, research has demonstrated that performing three acts of kindness per week for four consecutive weeks can reduce loneliness, with this effect continuing for two weeks after the study ends [18]. Similarly, in a three-week experiment, researchers found that the positive impact of prosocial acts on well-being persisted for two weeks post-study [42]. In light of these findings, it is possible that if people engage in prosocial behavior toward an agent over a longer period, their well-being may also experience sustained improvement. Future research could test this hypothesis.

Table 3: Summary of findings for the hypothesis and the research question.

| Hypothesis | Supported | Not Supported |
| --- | --- | --- |
| **H1:** Helping AI agents improves people's well-being, including (a) increased positive affect and (b) increased self-esteem, along with (c) reduced negative affect and (d) reduced loneliness, compared to those who do not help. | **H1(d):** Participants who helped the AI agent had a greater decrease in loneliness compared to those who did not help. | **H1(a), (b), & (c):** There were no significant differences between participants who helped the AI agent and those who didn't help in terms of positive affect, negative affect, and self-esteem. |
| **Research Question** | **Results** | |
| **RQ1:** How do AI agents satisfying human basic psychological needs (competence, autonomy, and relatedness) affect people's well-being when they help these agents? | **Competence**<br>- The competence group had a greater increase in positive affect and decrease in loneliness compared to the non-competence and the no-help groups, and also had a greater decrease in negative affect than the non-competence group.<br>- No significant differences in changes in self-esteem among the no-help, non-competence, and competence groups, nor were changes in negative affect between the competence and the no-help groups.<br>**Autonomy**<br>- The autonomy group had a greater increase in positive affect and a decrease in loneliness compared with the no-help and the non-autonomy groups, and also had a greater increase in self-esteem than the non-autonomy group.<br><br>- No significant differences in changes in negative affect among the no-help, the non-autonomy, and the autonomy groups, nor were in changes in self-esteem between the autonomy and the no-help groups.<br>**Relatedness**<br>- The non-relatedness group experienced a greater increase in positive affect and a marginally significant decrease in loneliness compared to the no-help group.<br>- No significant differences in changes in negative affect and self-esteem among the no-help, the non-relatedness, and the relatedness groups, nor were changes in positive affect, negative affect, self-esteem, and loneliness between the relatedness and the non-relatedness groups. | |
| **RQ2:** What is the interaction effect of AI agents satisfying different psychological needs on people's well-being when they help these agents? | - Participants without relatedness to the AI agent experienced greater increases in positive affect when their competence and autonomy needs were met, compared to those with relatedness.<br><br>- Participants with relatedness to the AI agent experienced a greater decrease in positive affect when their competence and autonomy needs were not met, compared to those without relatedness. | |

Based on our findings, we want to propose the concept of "Reciprocal AI." This envisions a human-AI collaboration where not only the AI provide help to humans, but it can also request help from humans—a reciprocal interaction that not only boosts human-AI collaboration outcomes but also enhances human well-being. Notably, while Okada has proposed the "weak robot" concept which suggests designing robots with imperfections that require human help to improve well-being [25], this idea primarily focuses on embodied agents and overlooks disembodied ones, which might be more ubiquitous due to their cost-effectiveness and flexibility. Besides, the weak robot concept advocates for intentionally designing robots with vulnerabilities, which might be valuable in scenarios aimed at increasing user engagement or enhancing users' mastery of a skill, such as allowing language learners to correct robots' English word errors to improve their grasp of the language [64]. However, in scenarios where the outcome of human-AI collaboration is critical, such as in human-AI clinical decision-making [34], intentionally weakening AI could impair the outcomes. Therefore, we suggest that in these scenarios, there might be no need to deliberately design weak AI agents. Following the trend of human-AI collaboration towards achieving complementary expertise [77], AI and humans may differ in expertise. Under the "Reciprocal AI" concept, AI in these scenarios could proactively request help in areas where it lacks, as enabling humans to help AI in these aspects might also enhance human well-being.

## 5.2 The Benefits of AI Satisfying Competence and Autonomy Needs When Humans Help AI

In line with SDT [22, 58], our study found that participants whose needs for competence and autonomy were satisfied by AI agents when helping the agents had a greater increase in positive affect and a decrease in loneliness, compared with those whose needs were not satisfied and those who did not help the agents. In the meanwhile, for those whose needs for competence and autonomy were not satisfied by the AI agents while they helped the agents, their positive affect and loneliness did not have a significant difference from those who did not help. Collectively, these findings uniquely define the boundary conditions where helping AI agents can enhance human well-being—that is, only when human needs for autonomy and competence are satisfied while helping AI agents, their well-being can be enhanced. Besides, these findings contribute to the understanding of the importance of AI agents satisfying human basic psychological needs in human-AI interactions for human well-being, as prior studies have primarily focused on how meeting these needs affects human engagement with tasks or agents [13, 38, 66]. In this light, when future research examines human well-being in human-AI interactions, it is encouraged to consider the role of AI in satisfying these human psychological needs.

Specifically, our findings showed that participants' well-being did not improve if they perceived their help as non-autonomous. Furthermore, not satisfying participants' needs for autonomy when they helped AI agents tended to have detrimental effects compared with not helping, as those whose needs for autonomy were satisfied had a greater increase in self-esteem compared only to those whose needs were not satisfied, not to those who did not help. These findings might be particularly insightful for situations where interaction with AI lacks voluntary choice. For example, reCAPTCHA implicitly involves humans helping AI label data [48], but since people must do this to access their desired content, it is likely that people perceive such help as non-autonomous, which might lead to people's decreased well-being. Also, as AI increasingly functions as a social actor and may even possess power in workplaces [21, 39], people helping AI under these situations might feel compelled by job requirements or power dynamics, which may lead to decreased well-being as well. When future research examines humans helping AI agents, it is thus important to pay attention to whether human needs for autonomy are satisfied during this process to determine the impact of helping on human well-being.

Furthermore, we manipulated autonomy based on the common definition of autonomy in prosocial behavior—whether a helper has a personal choice or volition in providing help [5, 70]. Specifically, we offered participants in the autonomy group the option to choose whether to help the agent, thus making their help more driven by intrinsic motivation, and required participants in the non-autonomy group to help the agent, so their help was more motivated by external pressure. However, this approach might lead to an issue: participants in the autonomy group could choose not to help the agent. In scenarios where it is crucial to obtain people's help for the agent, this may not align with the desired outcomes of stakeholders. The challenge, then, is finding ways to require people to help the agent while still giving them a sense of autonomy.

Previous research offers some potential solutions, such as giving people more control [76], like choosing whether to help through text, voice, or images, and enhancing transparency [53], such as clearly explaining how people's help will be used. More studies are encouraged to explore how to enhance people's sense of autonomy while still requiring them to help agents.

Additionally, we found that participants' well-being improved when their needs for competence were satisfied by AI agents during the helping process. This may help explain why people felt happy after helping robots, a phenomenon observed in previous studies but not empirically tested [50, 75]. Based on our findings, it might be because people experienced an enhanced sense of achievement from helping robots. Meanwhile, we also found that participants' well-being *only* improved when AI agents satisfied their needs for competence during the helping process, which indicates the importance of allowing people to sense the positive influence of their help offered to AI agents. Notably, in certain scenarios, the positive influence of helping intelligent agents may be more explicit, such that even without explicitly informing people of the positive impact, they may still gain a sense of achievement from helping. For instance, when helping Tweenbot—which clearly shows its destination on a flag and requests help to reach it—people can directly understand the positive impact of their help [28] and thus may have enhanced well-being out of helping. Conversely, in some scenarios, such positive impacts might be implicit, such as pointing out biases in a model's decisions [46]. People may not be able to clearly perceive how much the model's fairness has enhanced because of their help, thus they might be less likely to experience increased well-being. In these cases, it is important for stakeholders to consider providing clear feedback about the positive influence of users' help for better user outcomes.

## 5.3 AI Agents Satisfying Human Relatedness Needs When Humans Help AI

Contrary to SDT [22, 58], our study showed that relatedness with AI agents did not enhance participants' well-being by helping the agents. One possible explanation is that the relatedness formed between the participants and the AI agents did not reach the level of a strong tie, as previous research examining the well-being benefits of satisfying relatedness while helping has primarily focused on strong versus weak ties [72]. Specifically, the reason helping a strong tie might be more beneficial than helping a weak tie is that people are likely to interact with strong ties in the future [4], which makes prosocial behavior more advantageous for long-term relationships. To test this hypothesis, future research could establish strong ties between people and AI agents, such as enhancing self-disclosure on both sides over longer interactions [60], to examine the well-being effects of helping AI agents with whom people have established strong ties.

Another reason why relatedness with AI agents did not enhance participants' well-being by helping the agents could be because of the fundamental difference in relatedness established with humans and with AI. As indicated by previous research, people perceive relatedness with AI as more self-oriented and serving their own needs, whereas relatedness with humans is seen as more reciprocal, as human behavior is perceived as more voluntary [7]. Therefore, when people help other humans, they might feel that this prosocial act could further endear them to the other person. However, when helping AI, they might not hold this expectation, as the AI will maintain relatedness with them regardless of whether they provide help or not.

This explanation further highlights the potential difference between helping humans and helping AI, particularly that helping AI may be less rewarding than helping humans. One possible reason for such a difference, as mentioned above, is that when people help other humans, they may expect their actions to strengthen their relationship or earn the other person's affection [67]. However, such expectations may not exist when helping AI, which could result in fewer well-being benefits. Another possible reason is that helping other humans can enhance the helpers' social reputation, making the public more willing to cooperate with the helpers and offer help to them, a phenomenon known as indirect reciprocity [51]. However, when helping AI agents, people may not expect to gain social reputation benefits, which may lead to fewer well-being benefits than helping humans. In this process, people's perceptions of AI may also play a role [49]: the more people see AI agents as social actors, the more likely they are to unconsciously apply social scripts when helping AI, which may extend the positive effects of helping humans to helping AI.

In this light, one potential approach to bridge the potential gap between helping AI and helping humans, if any, is to increase the anthropomorphism of agents, as more human-like agents may lead people to unconsciously apply social scripts to their interactions [49] and thus may achieve similar well-being benefits from helping. Another potential approach is to design the development of relationships between humans and agents to be gradual, allowing people to anticipate a strengthening bond with the agent as a result of helping. Research could also explore whether helping AI agents can enhance a person's social reputation, as realizing this could make the benefits of helping AI more similar to those of helping humans. Finally, future research could investigate the role of AI as a mediator for helping, where people help AI, but their help indirectly benefits other humans. For example, prior work has shown that chatbots can act as mediators by asking for donations to environmental refugees and expressing gratitude for the help [47]. In such cases, although the help is directed at the AI, it may feel more rewarding because the ultimate beneficiaries are other humans.

In addition, our findings also highlight the positive effect of *not* satisfying participants' needs for relatedness on their positive affect when they helped AI agents. These findings might be explained by expectancy violation theory: when people experience outcomes better than expected, they perceive the outcomes more favorably, and when outcomes are worse than expected, they perceive them less favorably [8]. Based on this theory, it is likely that participants in our study who had no relatedness with the agent had lower expectations of the agent, thus when they later received support for their autonomy and competence, they perceived this as a positive violation of expectancy and had a more positive effect. Conversely, participants who had relatedness with the agent had higher expectations of the agent, thus not receiving subsequent autonomy and competence support was perceived as a negative violation and led to decreased positive affect. Since prior research has also emphasized the importance of meeting human expectations during human-AI interaction [20, 55], future research can further test our proposed explanations and, more broadly, explore how expectations for AI agents can be set or calibrated to improve user outcomes.

### 5.4 Design Implications

Our findings suggest that AI agents can be designed to ask for help and engage humans in helping behaviors to enhance human well-being. Such designs can manifest in two ways. The first can apply to everyday human-AI interactions or collaborations. Given that AI agents are encouraged to communicate uncertainty to people to improve transparency and trust [6], designers can prompt AI agents to ask for users' help after the agents express uncertainty, which may both contribute to agents' performance and enhance human well-being. The second way involves deliberately designing AI agents to be imperfect and prompting them to seek human help. Just as Nishiwaki et al. created a robot that struggles to hand out tissues and needs humans' help [50], designers can create AI agents that have difficulties with tasks like understanding human emotions or solving math problems, and then ask for help from humans to engage them in helping behaviors.

However, it should be noted that these imperfect AI agents might be better suited for scenarios where the outcome of the human-AI collaboration is not critical, such as scenarios outside of human-AI clinical decision-making [34], because a reduction in capability in these scenarios might impair outcomes and lead to adverse effects. Additionally, the degree of imperfection must be carefully balanced, as people might naturally expect AI agents to display intelligence, and continuous errors could decrease people's acceptance of them. To address this, a possible approach is to adjust people's expectations—preparing them for AI imperfections and enhancing their acceptance of such AI [29].

Additionally, an important ethical consideration for designers is that creating fake scenarios where AI agents appear to need help, but actually do not, may undermine people's trust and even negatively impact their well-being. As demonstrated by previous research, deceptive behaviors by intelligent agents can reduce people's trust, even when such behaviors are intended to benefit people [56]. To address this issue, AI agents can primarily request help in genuine scenarios where they truly require human help and avoid deception. As proposed by Mollick, the relationship between AI agents and humans is evolving into "co-intelligence," where AI acts as a cooperative partner rather than just an automated tool [43]. Within this relationship, designers can identify areas where AI agents lack capabilities and genuinely need human help, such as in morality and the humanities [43]. Soliciting help in these areas may not only improve people's well-being but also avoid the negative effects of deception.

During the process of humans helping AI agents, designers could also consider features that satisfy people's competence needs. To achieve this, designers can consider providing three elements employed in this study: an understanding of personal impact, explanations behind questions, and feedback on contributions. For understanding of personal impact, designers can make people feel important to the task when the agents ask for help, which can be achieved through personalization based on users' past interaction data. For example, in tasks that require people's help with image recognition [19], AI agents can inform people that their insights are valuable because their past performance in image recognition suggests they excel at it. Regarding explanations for questions, designers can prompt AI agents to explain why such questions are being asked to allow people to understand the potential impact of their help. Finally, for feedback on contributions, designers can let AI agents express how people's help makes a change to the agents after the agents receive help. As demonstrated in this study, the change does not need to be tangible. Even verbal acknowledgments suffice.

Another consideration is satisfying people's autonomy needs. One way to achieve this is to provide choices to people, a method employed in this study. This might be particularly important in workplaces where users might feel obliged to help others [37]. In such a setting, AI agents could provide people with the choice to help, so that people can perceive their help as more voluntary and thus derive well-being benefits through helping. Beyond choice, designers can also consider other proven methods such as avoiding terms like "should" or "must", instead conveying flexibility with words like "could" or "might" [63], or having agents explain the importance of the helping task to make people feel that their help is meaningful rather than pressured [63]. Nonetheless, it should be noted that these autonomy-enhancing techniques have not been validated in this study. Future research can investigate if these techniques indeed enhance people's sense of autonomy and well-being when they help AI agents.

Finally, although our study did not find that satisfying participants' relatedness needs through AI agents led to enhanced well-being, we discovered that when participants' needs for relatedness were satisfied but needs for autonomy and competence were not, participants' positive affect decreased. In this light, designers might be cautious that when users perceive relatedness with AI agents, their needs for autonomy and competence are also addressed to avoid potential detrimental effects on users' positive affect. This might be especially important for social chatbots that tend to develop relatedness with humans [52, 74], such as personal assistants Alexa, mental health and elderly care attendants Woebot, and friendship companions Replika. For instance, when users offer emotional support to Replika [54], it is important to let Replika affirm the value of users' support to prevent negative expectation violations and to safeguard users' well-being.

### 5.5 Limitation and Future Directions

Firstly, the helping task involved in this study was to help AI agents with messaging app development. However, there exist other helping tasks, such as helping AI with image recognition [19], improving AI fairness [46], and providing emotional support to AI [54]. Future research could explore the impact of helping AI agents with these additional tasks to test the generalizability of our findings.

Secondly, this study did not explore the effects of AI agents asking for help on people's perceptions of the agents. Although research has found that expressing uncertainty can enhance people's trust toward AI [6], it has also been found that AI agents requesting help amplify people's perceptions of the agents' incompetence and reduce trust [46]. In light of this discrepancy, future research can examine how AI agents requesting help affect people's perceptions of the agents.

Thirdly, this study did not explore why some people are willing to help AI agents while others are not. Although prior work has investigated factors influencing people's willingness to help robots, such as the robot's politeness strategies, social status, size of the request, and people's familiarity with robots [61], factors influencing people's willingness to help disembodied agents remain underexplored. Future research is encouraged to explore this issue.

Fourthly, this study did not examine whether the observed well-being benefits persist over time. Given that prior longitudinal studies have shown that helping others can enhance people's more stable well-being such as life satisfaction [9], future longitudinal studies are needed to determine the long-term well-being impact of helping AI agents.

Finally, future research could explore factors that influence the well-being benefits derived from helping AI agents. Specifically, the role of individual differences, such as personality traits or prior attitudes towards AI, could be explored to understand who benefits most from these interactions. Additionally, it would be interesting to investigate how different levels of help from humans affect their well-being. This could inform the optimal help-seeking behavior from AI agents that effectively engages users and enhances their well-being.

## 6 Conclusion

Although helping others has long been found to improve human well-being, it remains unclear whether this effect extends to helping AI agents. This study showed that helping AI agents could reduce participants' loneliness. When participants' needs for competence and autonomy were satisfied by AI agents while helping these agents, there was a further increase in positive affect and a decrease in loneliness. However, satisfying participants' relatedness needs did not impact their well-being. Collectively, these findings provide a new perspective for research on AI for well-being and offer guidance for designing AI agents so that humans can derive well-being benefits from helping these agents.

## Acknowledgments

This research was funded by the Yale-NUS Seed Grant (A-8001353-00-00) and the Centre for Computational Social Science and Humanities, National University of Singapore (A-8002954-00-00).